\documentclass[aps,pre,twocolumn,showpacs,amsmath,amssymb, longbibliography,notitlepage,superscriptaddress]{revtex4-2} 

\usepackage{xcolor}
\usepackage{graphicx}
\usepackage{siunitx}
\usepackage{placeins}
\usepackage{ulem}

\definecolor{darkblue}{rgb}{0,0,0.6}
\definecolor{darkred}{rgb}{0.6,0,0}
\usepackage[colorlinks=true,urlcolor=darkblue, citecolor=darkblue,linkcolor=darkred,hyperfootnotes=false]{hyperref}

\newcommand{\dk}{\Delta\kappa}

\newcommand{\srr}{\sigma_{rr}}
\newcommand{\srt}{\sigma_{r\theta}}
\newcommand{\stt}{\sigma_{\theta\theta}}
\newcommand{\hstt}{\hat\sigma_{\theta\theta}}

\newcommand{\ind}[1]{_\mathrm{#1}}

\begin{document}

\title{Shape of a membrane on a liquid interface with arbitrary curvatures}

\author{Zachariah S. Schrecengost}
 \affiliation{Department of Physics, Syracuse University, Syracuse, NY 13244}
 \affiliation{BioInspired Institute, Syracuse University, Syracuse, NY 13244}

\author{Seif Hejazine}
 \affiliation{Department of Physics, Syracuse University, Syracuse, NY 13244}
 \affiliation{Department of Electrical Engineering and Computer Science, Syracuse University, Syracuse, NY 13244}

\author{Jordan V. Barrett}
 \affiliation{Department of Physics, Syracuse University, Syracuse, NY 13244}
 
\author{Vincent D\'emery}
 \email{vincent.demery@espci.psl.eu}
 \affiliation{Gulliver, CNRS, ESPCI Paris, PSL Research University, 10 rue Vauquelin, 75005 Paris, France}
 \affiliation{Univ Lyon, ENS de Lyon, Univ Claude Bernard Lyon 1, CNRS, Laboratoire de Physique, F-69342 Lyon, France}

\author{Joseph D. Paulsen}
 \email{jdpaulse@syr.edu}
 \affiliation{Department of Physics, Syracuse University, Syracuse, NY 13244}
 \affiliation{BioInspired Institute, Syracuse University, Syracuse, NY 13244}

\begin{abstract}
We study the deformation of a liquid interface with arbitrary principal curvatures by a flat circular sheet. 
Working first at small slopes, we determine the shape of the sheet analytically in the membrane limit, where the sheet is inextensible yet free to bend and compress. 
We find that the sheet takes a cylindrical shape on interfaces with negative Gaussian curvature. 
On interfaces with positive Gaussian curvature, an inner region still adopts a cylindrical shape while the outer region is under azimuthal compression.
Numerical energy minimization confirm our predictions and show that this behavior holds for finite slopes.
Experiments on a thin polystyrene film at an anisotropic air-water interface show consistent behaviors.
\end{abstract}

\maketitle

Experiments where a thin elastic solid is placed on a surface with a different metric can reveal how a physical system grapples with the basic mathematical conflict of two incompatible metrics~\cite{Marder2007Crumpling,Paulsen2019}. 
One avenue for studying these problems has been to stamp~\cite{Hure2012, Davidovitch2019} or adhere~\cite{Yao2013, Bense2020, Box2023Delamination} a sheet onto a curved substrate, thereby forcing the sheet to adopt a wrinkle pattern that approximates the confining geometry. 
In some cases, a liquid surface can provide such strong confinement, for instance when a thin curved shell is laid on a flat liquid bath~\cite{Aharoni2017,Tobasco2021CurvatureDriven, Tobasco2022Exact}. 
In other cases, a shell can impose its own metric on a droplet~\cite{Timounay2021Sculpting}, a response termed ``sculpting''. 
Lying between these two extremes, both the liquid and the sheet can deform into a non-trivial shape that matches neither of the original geometries~\cite{King2012, Paulsen2015, Bae2015Measuring, Kumar2018}. 

\begin{figure}
  \begin{center}
    \includegraphics[scale=1.]{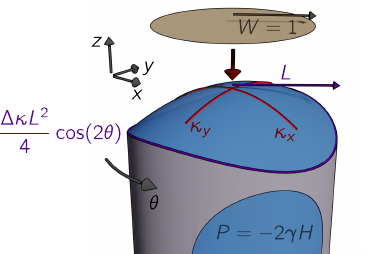}
    \caption{Setup: a flat circular sheet with radius $W=1$ is placed on a liquid interface with principal curvatures $\kappa_x$ and $\kappa_y$. The curvature difference $\dk=\kappa_x-\kappa_y$ is set by the shape of the edge while the mean curvature $H=(\kappa_x+\kappa_y)/2$ is set by the pressure applied under the interface.}
    \label{fig:setup}
  \end{center}
\end{figure}

While much has been elucidated about the behavior of a thin sheet confined to a liquid interface, 
the vast majority of these studies were carried out in axisymmetric configurations, leaving the case with different principal curvatures $\kappa_x$ and $\kappa_y$, as shown in Fig.~\ref{fig:setup}, unexplored.
For the case of solid confinement, previous work  has focused on the role of the Gaussian curvature, $K=\kappa_x\kappa_y$, at driving the film response~\cite{Hure2012,Yao2013}. 
For instance, in stamping or adhesion, the deformation of the sheet is primarily controlled by the difference between the Gaussian curvature of the sheet $K\ind{sh}$ and that of the solid substrate, $K\ind{sub}$ \cite{Hure2012,Hohlfeld2015,Davidovitch2019, Box2023Delamination}. 
This result extends to a shell confined to a flat liquid, as gravity prevents large deformations~\cite{Aharoni2017, Timounay2021Sculpting, Tobasco2021CurvatureDriven, Tobasco2022Exact}. 
However, whether this result generally applies to liquid interfaces remains unclear. 

\begin{figure}
  \begin{center}
    \includegraphics[scale=.9]{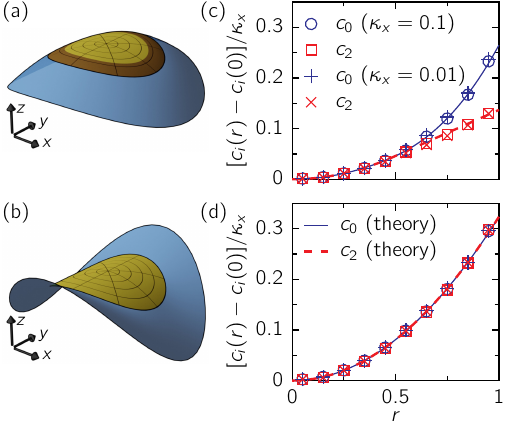}
    \caption{(a,b) Rendering of the shapes predicted for $\kappa_y=\kappa_x/2$ and $\kappa_y=-\kappa_x/2$, respectively. The shape is cylindrical in the yellow region.
    (c, d) values of $c_0(r)$ and $c_2(r)$ obtained from the fits (symbols) and theoretical predictions (solid lines) for $\kappa_y=\kappa_x/2$ and $\kappa_y=-\kappa_x/2$, respectively, for $L=1.2$.
    }
    \label{fig:theory}
  \end{center}
\end{figure}

Here, we study this problem using analytical calculations, numerical energy minimization and experiments.
We use the simplest theoretical model: the sheet is treated as an inextensible, highly flexible membrane that can compress and wrinkle freely~\cite{Paulsen2015,Paulsen2017}.
In the limit of small slopes, we determine the shape of the sheet analytically.
For both positive and negative Gaussian curvatures, we find a region of the sheet that forms a cylindrical shape, where the sheet only curves along a single principle axis [Fig.~\ref{fig:theory}(a,b)]. 
We find that the deformations are not controlled solely by the Gaussian curvature, but instead by the mean curvature and the difference between $\kappa_x$ and $\kappa_y$. 
Our predictions are confirmed and extended to finite slopes by numerical energy minimization and are compatible with the experimental observation of a thin polystyrene film placed on an anisotropic
air-water interface.
Our results comprise a fundamentally different response from related studies with other forms of confinement~\cite{Hure2012, Yao2013, Hohlfeld2015, Aharoni2017, Davidovitch2019, Tobasco2021CurvatureDriven, Tobasco2022Exact, Box2023Delamination}. 

A liquid interface with arbitrary curvatures can be formed by applying a pressure $P=-2\gamma H$ under the interface, where $\gamma$ is the surface tension of the interface, and pinning its edge at a frame with radius $L$ and height $\zeta(L,\theta)=\dk L^2\cos(2\theta)/4+HL^2/2$ (Fig.~\ref{fig:setup}).
We assume that the vertical extension of the interface, scaling as $\kappa_i L^2$, is small with respect to the capillary length, so that the effect of gravity is negligible.
Last, we restrict ourselves to small slopes.
Under these conditions, the height $\zeta$ of the liquid interface, which satisfies $\nabla^2\zeta=2H$ and Dirichlet boundary conditions at the frame reads, in polar coordinates:
\begin{equation}
    \zeta_0(r,\theta)=\frac{H}{2}r^2+\frac{\dk}{4}r^2\cos(2\theta),
\end{equation}
where $H$ and $\dk$ are related to the principal curvatures $\kappa_x$ and $\kappa_y$ by $H=(\kappa_x+\kappa_y)/2$ and $\dk=\kappa_x-\kappa_y$.
We choose $x$ as the most curved direction, $\kappa_x>|\kappa_y|$.
While all the calculations below can be performed for arbitrary values of $L$, for simplicity we give in the main text only the results for $L\to\infty$~\cite{SM}.

We then place a circular inextensible sheet with radius $W$ at the center of the interface (we now use $W$ as the unit length). 
The sheet is treated as inextensible but with zero bending modulus; such a membrane strongly resists in-plane stretching while being free to wrinkle and deform under minute compressive stresses; such small-scale wrinkling can allow lengths in the sheet to effectively shorten \cite{Pak10,Paulsen2015,Paulsen2017,SM}.
We assume a vertical displacement of the sheet ($r\leq 1$) of the form
\begin{equation}\label{eq:vertical_displacement}
    \zeta(r,\theta) = c_2(r)\cos(2\theta)+c_0(r).
\end{equation}
With the sheet, the shape of the liquid interface ($r\geq 1$) is perturbed and takes the form
\begin{equation}\label{eq:liquid_height_gen}
    \zeta(r,\theta)=\frac{H}{2}r^2+\left[\frac{\dk}{4}\left(r^2-\frac{1}{r^2}\right)+\frac{c_2(1)}{r^2}\right]\cos(2\theta).
\end{equation}

We use force balance to determine the shape of the sheet.
We start with the case where the edge of the sheet is under azimuthal compression so that the azimuthal component of the stress vanishes.
The in-plane force balance gives $\partial_r(r\srr)=0$, where $\srr$ is the radial component of the stress, hence $\srr(r)=\gamma/r$ due to the boundary condition $\srr(1)=\gamma$.

We now use the vertical force balance: $\srr(r)\partial_r^2\zeta(r,\theta)=2\gamma H$.
Inserting the vertical displacement (\ref{eq:vertical_displacement}) and the radial stress $\srr(r)=\gamma/r$, we find $c_2''(r)=0$ and $c_0''(r)=2Hr$.
Using the continuity of the height and slope at the edge of the sheet leads to
\begin{align}
    c_0'(r) & = Hr^2,\\
    c_2(r)  & = \left[\dk-2c_2(1)\right]r+3c_2(1)-\dk. \label{eq:c2}
\end{align}
Unless $\dk=c_2(1)=0$, these expressions cannot hold down to the center of the sheet as there would be a singularity at the origin.
The slope along $y$, $c_0'(r)-c_2'(r)$ reaches $0$ at $r=\ell$ where $H\ell^2=\dk-2c_2(1)$, pointing to a cylindrical shape.
We thus assume that the shape is cylindrical for $r\leq \ell$, meaning that $c_0'(r)=c_2'(r)=H\ell r$: the sheet is curved only in the $x$ direction, with curvature $\kappa_c=2H\ell$.
Matching the height of the cylinder at $r=\ell$ with Eq.~(\ref{eq:c2}) leads to $c_2(1)=(\dk-H\ell^2)/2$ and
\begin{equation}
    \ell^2\left(3-\ell\right) = \frac{\dk}{H} = \frac{2(\kappa_x-\kappa_y)}{\kappa_x+\kappa_y}.
\end{equation}
This equation admits a solution in the range $[0,1]$ for $\dk/H$ in the range $[0,2]$, corresponding to $\kappa_y$ in the range $[0,\kappa_x]$.
As expected, $\ell=0$ in the axisymmetric case $\kappa_x=\kappa_y$ and $\ell=1$ in the cylindrical case $\kappa_y=0$.
A representation of the predicted shape for $0 < \kappa_y < \kappa_x$ is shown in Fig.~\ref{fig:theory}(a).


Based on the calculation above, we assume that the shape remains cylindrical for $\kappa_y<0$: $c_0(r)=\frac{1}{4}\kappa_c r^2+c_0(0)$, $c_2(r)=\kappa_c r^2/4$.
As the edge is under azimuthal tension, it can carry a stress; moreover, since the sheet is inextensible, the stress can be localized at the edge, and we denote this singular part $\hstt(r)=\Sigma\delta(r-1)$.
This singular azimuthal stress can give rise to a stress jump in the longitudinal direction, as can be seen from the in-plane force balance $\partial_r[r\srr(r)]=\stt(r)$, leading to
\begin{equation}\label{eq:ipfb_edge}
    \srr(1^-) = \gamma-\Sigma.
\end{equation}
The singular azimuthal stress also allows a slope discontinuity at the edge, which is given by the vertical force balance:
\begin{equation}\label{eq:oopfb_edge}
    \gamma\partial_r\zeta(1^+,\theta) -\srr(1^-)\partial_r\zeta(1^-,\theta)+\Sigma\partial_\theta^2\zeta(1,\theta)=0.
\end{equation}
Combining these equations with Eq.~(\ref{eq:liquid_height_gen}), we can determine the shape of the sheet and the singular edge stress:
\begin{align}
  \kappa_c & = \frac{2}{3}(H+\dk),\label{eq:C_k2neg}\\
  \frac{\Sigma}{\gamma} & = 1-\frac{2H}{\kappa_c}=1-\frac{3H}{H+\dk}.
\end{align}
The edge stress should be positive, which is the case for $2H\leq \dk$, or $\kappa_y\leq 0$, as expected.
A representation of the predicted shape for $-\kappa_x < \kappa_y < 0$ is shown in Fig.~\ref{fig:theory}(b).


We compare our theoretical predictions to numerical energy minimization using ``Surface Evolver''~\cite{Brakke1992, SM}.
We allow mesh edges within the sheet to shorten at zero cost, to capture the effect of small-amplitude, short-wavelength wrinkles that can form in the sheet. 
The sheet nevertheless resists stretching, which we implement using a large stretching modulus, $Y = 10^{12} \gamma$. 
First, we check that the shape is of the form (\ref{eq:vertical_displacement}) by plotting the vertical displacement over a narrow annulus; the functions $c_i(r)$ are then obtained by fitting the angular dependence at different radii~\cite{SM}.
The results are compared to the predictions for several values of $\kappa_x$ and $\kappa_y=\pm\kappa_x/2$ in Fig.~\ref{fig:theory}(c,d); a very good agreement is obtained.

Symmetry allows to deduce the shape of the sheet for arbitrary curvatures from the calculations in the case where $\kappa_x < |\kappa_y|$.
There are two situations where $\kappa_x=|\kappa_y|$.
In the axisymmetric configuration, $\kappa_x=\kappa_y$, the solution is the parachute shape predicted by Taylor~\cite{Taylor1919}.
The zero mean curvature configuration, $\kappa_x=-\kappa_y$, deserves more attention.
Our prediction is that for negative Gaussian curvatures the sheet adopts a cylindrical shape, which is flat in the less curved direction of the interface.
As a consequence, the flat direction changes abruptly when $\kappa_y$ crosses $-\kappa_x$.
At the transition, because the pressure across the interface is zero, the sheet is floppy and many configurations minimize the energy.

We have constructed solutions that satisfy force balance and the boundary conditions~\cite{SM}.
Uniqueness theorems for such traction boundary value problem require the elasticity to be positive definite~\cite{Knops1971Uniqueness}, which is not the case in our model because the sheet can compress under zero compressive stress.
Nevertheless, the agreement with the numerical simulations allows us to conjecture that the solution is unique for $H\neq 0$.
We also note that, because in our model the bending modulus is zero, there are no continuity constraints on the second and third derivatives of the height.

\begin{figure}
  \begin{center}
    \includegraphics[scale=.9]{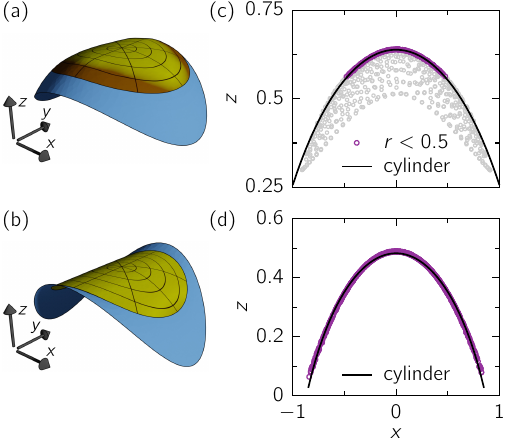}
    \caption{(a,b) Renderings of the shapes obtained by numerical energy minimization and (c,d) positions of the vertices projected on the $(x,z)$ plane (circles) and fit with a cylinder (solid line).
    The parameters used are $L=1.2$ and (a,c) $H=-0.7$, $\Delta\kappa=-1$ and (b,d) $H=-0.4$, $\Delta\kappa=-1.6$.
    In (a,b), the color indicates the distance to the cylinder fitted in (c,d), it is yellow for distances less than $0.04$.
    In (c), the cylinder is fitted only on the points at a distance $r<0.5$ from the center, which are shown in purple.}
    \label{fig:finite}
  \end{center}
\end{figure}

We determine the shape of the sheet in presence of finite slopes numerically.
We find that the sheet can adopt a cylindrical shape over a central region or over the whole sheet, depending on the pressure applied under the interface and the shape of the frame (Fig.~\ref{fig:finite}).
Finding analytical solutions for finite slopes is a considerable challenge for two main reasons.
First, there is no interface with constant mean and Gaussian curvatures, so that the very question to ask is different.
Second, even in the simplest, axisymmetric situation where $\kappa_x=\kappa_y$, the sheet breaks axisymmetry and no analytical solution is known~\cite{Paulsen2015}.

\begin{figure*}
  \begin{center}
    \includegraphics[width=\linewidth]{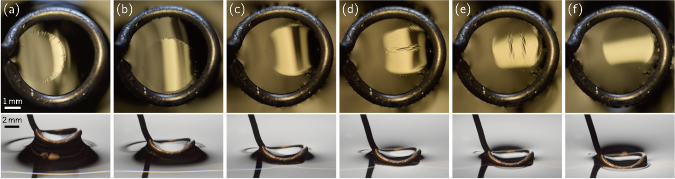}
    \caption{Top and side views of a circular polystyrene film of thickness $\qty{107}{\nano\meter}$ floating on water. 
The water meniscus is pinned to an undulating metal frame that we raise and lower to vary the pressure across the interface. 
In (a), interface has positive Gaussian curvature and the sheet forms radial wrinkles. 
Lowering the frame, the wrinkles eventually vanish (b) and the sheet remains absent of visible wrinkles over a finite interval in the frame height up to (c). 
In (d), the film buckles in its interior; the buckles change orientation in (e), marking a change in the axis of the gross cylindrical shape. 
The buckles vanish upon further lowering (f). Distance the metal frame was lowered between neighboring images in (a-f), respectively: $\qty{1.4}{\milli\meter}$, $\qty{0.9}{\milli\meter}$, $\qty{0.4}{\milli\meter}$, $\qty{0.3}{\milli\meter}$, $\qty{0.5}{\milli\meter}$.}
    \label{fig:exp}
  \end{center}
\end{figure*}

Finally, we test these predictions by realizing these boundary conditions in an experiment. 
We spin-coat polystyrene films of thickness $t$ and Young's modulus $E=\qty{3.4}{\giga\pascal}$, which we then cut into a disc of radius $\qty{1.7}{\milli\meter}$ and float onto a dish of deionized water ($\gamma = \qty{72}{\milli\newton\per\meter}$)~\cite{Huang2007,Timounay2020, SM}.
To impose a general surface geometry around the film, we take a piece of $\qty{0.8}{\milli\meter}$ stainless steel wire and bend its end into a circle, and then deform the circle into an undulating frame with peak-to-peak height $\qty{0.74}{\milli\meter}$ and inner radius $\qty{2.4}{\milli\meter}$. 
We lower the frame onto the water surface to corral the film within it. 
Raising and lowering the frame via a micrometer stage allows to tune the pressure across the interface, via the hydrostatic pressure. 

Figure~\ref{fig:exp} shows the response of a film with $t=\qty{107}{\nano\meter}$. 
Starting at a large negative pressure [Fig.~\ref{fig:exp}(a)], we observe radial wrinkles at the edge of the film, which are expected when the interface has a positive Gaussian curvature. 
Lowering the frame, the wrinkles shorten and disappear entirely; the film remains completely absent of visible wrinkling over a finite interval in frame height [Fig.~\ref{fig:exp}(b-c)], pointing to a strain-free state, such as the cylindrical shape. 
Lowering the frame further, crumples \cite{King2012,Timounay2020} appear within the interior of the film [Fig.~\ref{fig:exp}(d)].
We attribute these crumples to the effect of gravity, which results in a nonuniform pressure accross the interface; this effect has been quantified for a shell on an axisymmetric interface~\cite{Timounay2021Sculpting}.
At a lower height still, these crumples change orientation, indicating that the axis of cylindrical deformation has changed [Fig.~\ref{fig:exp}(e)], in accordance with our predictions. 
These crumples disappear upon further lowering [Fig.~\ref{fig:exp}(f)], and the progression of patterns continues in reverse order (not shown). 
Finally, we repeated the experiment with additional films of thickness $t=\qty{49}{\nano\meter}$ and $\qty{111}{\nano\meter}$ and with a second wire frame, and we observed the same qualitative response.


Using theory, numerics, and experiment, we have found that a thin elastic sheet adopts a cylindrical shape when placed on a liquid interface with negative Gaussian curvature. 
In this situation, the sheet retains its metric and imposes it to the interface: it ``sculpts'' the interface, as a thin shell placed on a curved interface~\cite{Timounay2021Sculpting}.
There is a slight difference between the two cases: here, the sheet retains its metric but not its shape, contrary to the situation in Ref.~\cite{Timounay2021Sculpting}, so that the sculpting is weaker here.
This is due to the fact that there are more embeddings of the flat metric compared to the spherical one~\cite{Han2006Isometric}.
When the Gaussian curvature of the interface is positive, the sheet still adopts a cylindrical shape in its center, and thus partially sculpts the interface.
The ``rim'' that connects the sculpted region to the liquid interface is here purely geometric, while the one observed in Ref.~\cite{Timounay2021Sculpting} is due to the finite extensibility of the sheet.
Finally, we note that whether the sheet is flat or spherical, sculpting may occur when the Gaussian curvature of the sheet is larger than that of the interface.
Hence, while the precise response of the sheet does not depend solely on the Gaussian curvature mismatch, some aspects of the
sheet-interface interaction seem to be determined by the sign of the mismatch.

\begin{acknowledgments}
We thank Hillel Aharoni for fruitful discussions and Mokthar Adda-Bedia for useful comments on the manuscript.
Funding support from NSF-DMR-2318680 is gratefully acknowledged. 
\end{acknowledgments}

\appendix

\section{General and detailed derivation}
\label{sec:fb}

In this section, we present in more detail the derivation sketched in the main text, and generalize it to a finite radius $L$ of the interface.

\subsection{General shape of the sheet and the liquid interface}

As in the main text, we assume a vertical displacement of the sheet ($r\leq 1$) of the form
\begin{equation}\label{eq:vertical_displacement}
    \zeta(r,\theta) = c_2(r)\cos(2\theta)+c_0(r).
\end{equation}

With the sheet, the shape of the liquid interface ($r\geq 1$) is perturbed and takes the form
\begin{equation}\label{eq:liquid_height_gen}
    \zeta(r,\theta)=\frac{H}{2}r^2+\left(ar^2+\frac{b}{r^2}\right)\cos(2\theta),
\end{equation}
where the coefficients $a$ and $b$ are obtained by matching height at $r=1$ and $r=L$, leading to
\begin{align}
    a & = \frac{\frac{1}{4}\dk - Mc_2(1)}{1-M},\label{eq:liquid_a}\\
    b & = \frac{c_2(1)-\frac{1}{4}\dk}{1-M},\label{eq:liquid_b}
\end{align}
where we have defined
\begin{equation}
  M=L^{-4}.
\end{equation}

\subsection{Positive Gaussian curvature}

We start by recalling the in-plane and vertical force balance equations.
For zero bending modulus ($B=0$) and in the absence of shear stress ($\srt=0$) they read, respectively~\cite{King2012},
\begin{align}
  \partial_r(r\srr) & = \stt,\label{eq:ipfb}\\
  \srr\partial_r^2\zeta+\frac{1}{r^2}\stt\left(\partial_\theta^2\zeta+r\partial_r\zeta\right) & = -P. \label{eq:oopfb}
\end{align}
In these equations $\srr$ and $\stt$ denote the radial and azimuthal components of the stress, respectively.

We first consider the case where the edge of the sheet is under radial tension and azimuthal compression.
The radial component of the stress $\srr$ is nonzero and its azimuthal component vanishes, $\stt=0$.
The in-plane force balance (Eq.~(\ref{eq:ipfb})) reduces to $\partial_r(r\srr)=0$, so that $\srr(r)=\gamma/r$ due to the boundary condition at the edge of the sheet, $\srr(1)=\gamma$.

\begin{figure}
  \begin{center}
    \includegraphics[scale=1]{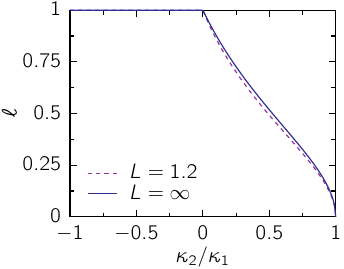}
    \caption{Size of the cylindrical region $\ell$ as a function of the curvature ratio $\kappa_2/\kappa_1$, in the infinite size limit (solid line) and for $L=1.2$ (dashed line, corresponding to $M\simeq 0.5$).
    }
    \label{fig:l}
  \end{center}
\end{figure}

We then use the vertical force balance (Eq.~(\ref{eq:oopfb})):
\begin{equation}
  \srr(r)\partial_r^2\zeta(r,\theta)=2\gamma H.
\end{equation}
Using the vertical displacement (\ref{eq:vertical_displacement}) and the radial stress $\srr(r)=\gamma/r$, it reads $\partial_r^2\zeta(r,\theta)=c_2''(r)\cos(2\theta)+c_0''(r)=2Hr$,
leading to $c_2''(r)=0$ and $c_0''(r)=2Hr$.
Moreover, the height and the slope should be continuous at the edge of the sheet, meaning that $\zeta(1,\theta)=c_2(1)\cos(2\theta)+c_0(1)=(a+b)\cos(2\theta)+\frac{H}{2}(1-L^2)$, and $\partial_r\zeta(1,\theta)=c_2'(1)\cos(2\theta)+c_0'(1)=2(a-b)\cos(2\theta)+H$, where $a$ and $b$ are given by Eqs.~(\ref{eq:liquid_a}, \ref{eq:liquid_b}).
Together, this leads to
\begin{align}
    c_0'(r) & = Hr^2,\\
    c_2(r)  & = 2(a-b)r-a+3b. \label{eq:c2}
\end{align}
Unless $a=b=0$, these expressions cannot hold down to the center of the sheet as there would be a singularity at the origin.
The slope along $y$, $c_0'(r)-c_2'(r)$ reaches $0$ at $r=\ell$ where $H\ell^2=2(a-b)$, pointing to a cylindrical shape.
We thus assume that the shape is cylindrical for $r\leq \ell$, meaning that $c_0'(r)=H\ell r$ and $c_2(r)=H\ell r^2/2$: the sheet is curved only in the $x$ direction, with curvature $\kappa=2H\ell$.
Matching the height of the cylinder at $r=\ell$ with Eq.~(\ref{eq:c2}) leads to
\begin{align}
  c_2(1) & = \frac{\dk-(1-M)H\ell^2}{2(1+M)},\\
  \frac{\dk}{H} & = \ell^2\left[3-\ell+M(1-\ell)\right]. \label{eq:l}
\end{align}
The equation (\ref{eq:l}) admits a solution in the range $[0,1]$ for $\dk/H$ in the range $[0,2]$, corresponding to $\kappa_y$ in the range $[0,\kappa_x]$ (Fig.~\ref{fig:l}).
We find that the effect of a finite size ($M>0$) is very weak.

We note that Eq.~(\ref{eq:l}) is a cubic equation, which can be solved analytically. 
When $M=0$, its solution reads
\begin{equation}
  \ell=1+2\cos\left(\frac{2\pi}{3}-\frac{1}{3}\arccos\left(\frac{2\kappa_2}{\kappa_1+\kappa_2}\right)\right).
\end{equation}

Finally, we show that force balance is satisfied in the cylindrical part.
To balance the pressure, the stress should be $\sigma_{xx}=2\gamma H/\kappa_c=\gamma/\ell$, which is indeed the value of the stress at $r=\ell$ in the outer part of the sheet. 
The boundary condition for the stress also imposes $\sigma_{yy}=\gamma/\ell$: for $r\leq\ell$, the sheet is under tension in all directions with an isotropic stress $\gamma/\ell$.

%
%
%

\subsection{Negative Gaussian curvature}

Based on the calculation above, we assume that the shape remains cylindrical for $\kappa_y<0$:
\begin{align}
  c_0(r)&=\frac{\kappa_c}{4}r^2+c_0(0),\\
  c_2(r)&=\frac{\kappa_c}{4}r^2.
\end{align}
This sets $c_2(1)=\kappa_c/4$, so that the coefficients for the shape of the interface are given by
\begin{align}
  a&=\frac{\Delta\kappa-M\kappa_c}{4(1-M)},\label{eq:a_neg}\\
  b&=\frac{\kappa_c-\Delta\kappa}{4(1-M)}.\label{eq:b_neg}
\end{align}

The edge being under azimuthal tension, it can carry a stress; moreover, since the sheet is inextensible, the stress can be localized at the edge, we denote this singular part $\hstt(r)=\Sigma\delta(r-1)$.
This singular azimuthal stress can give rise to a stress jump in the longitudinal direction, as can be seen from the in-plane force balance $\partial_r[r\srr(r)]=\stt(r)$ (Eq.~(\ref{eq:ipfb})), leading to
\begin{equation}\label{eq:ipfb_edge}
    \srr(1^-) = \gamma-\Sigma.
\end{equation}

The singular azimuthal stress also allows a slope discontinuity at the edge, which can be obtained from the vertical force balance (Eq.~(\ref{eq:oopfb})).
Multiplying Eq.~(\ref{eq:oopfb}) by $r$ and expanding, we obtain
\begin{equation}
  r\srr\partial_r^2\zeta+\stt \partial_r\zeta+\frac{1}{r}\stt\partial_\theta^2\zeta = -Pr.
\end{equation}
Now using the in-plane force balance (\ref{eq:ipfb}) in the second term, it reads
\begin{align}
  -Pr &= r\srr\partial_r^2\zeta+\partial_r(r\srr) \partial_r\zeta+\frac{1}{r}\stt\partial_\theta^2\zeta  \\
  &= \partial_r\left(r\srr\partial_r\zeta\right)+\frac{1}{r}\stt\partial_\theta^2\zeta.
\end{align}
Integrating the last relation between $1^-$ and $1^+$ leads to
\begin{equation}\label{eq:oopfb_edge}
  \gamma\partial_r\zeta(1^+,\theta)-\srr(1^-)\partial_r\zeta(1^-,\theta)+\Sigma\partial_\theta^2\zeta(1,\theta)=0.
\end{equation}

With the assumed shape for the sheet, we have $\partial_r\zeta(1^-,\theta)=\frac{\kappa_c}{2}[1+\cos(2\theta)]$ and $\partial_\theta^2\zeta(1,\theta)=-\kappa_c\cos(2\theta)$.
For the liquid interface, $\partial_r\zeta(1^+,\theta)=H+2(a-b)\cos(2\theta)$.
Separating the $\theta$ independent and $\cos(2\theta)$ contributions in the vertical force balance (\ref{eq:oopfb_edge}), we get
\begin{align}
    \gamma H - (\gamma-\Sigma)\frac{\kappa_c}{2} & = 0,\\
    2\gamma(a-b) - (\gamma+\Sigma)\frac{\kappa_c}{2} & = 0.
\end{align}

Combining these equations with Eqs.~(\ref{eq:a_neg}, \ref{eq:b_neg}), we can determine the shape of the sheet and the singular edge stress:
\begin{align}
  \kappa_c & = \frac{2[(1-M)H+\dk]}{3-M}.\label{eq:C_k2neg}\\
  \frac{\Sigma}{\gamma} & = 1-\frac{2H}{\kappa_c}
  =\frac{\dk-2H}{\dk+(1-M)H}.
\end{align}
The edge stress should be positive, which is the case for $2H\leq \dk$, or $\kappa_y\leq 0$, as expected.

The solution that we have obtained satisfies the vertical force balance: the sheet is cylindrical with curvature $\kappa_c$ and a uniform stress $\sigma=\gamma-\Sigma$, which satisfy $\kappa_c\sigma=2\gamma H$.

\subsection{Discussion of the membrane limit}

We quickly discuss here the implications of leaving the membrane limit.
The membrane limit assumes that the sheet is inextensible yet free to bend. 
Imparting a finite cost to bending would make the problem much more complicated by bringing in an energy scale associated with the curvature of the sheet.
Such resistance to bending is relevant for ``thick'' sheets, of the order of tens of micrometers~\cite{Py2007}.
On the other hand, the small but finite extensibility of ``thin'' sheets, of the order of a hundred nanometers, is clearly visible in experiments where a flat sheet is placed on a curved droplet~\cite{King2012} and where a curved shell is placed on a flat interface~\cite{Timounay2021Sculpting}. 
The sheets in these experiments would be fully wrinkled if they were inextensible; instead they display some unwrinkled regions where the sheet is stretched. 
For positive Gaussian curvatures, the extent of the unwrinkled region in the center of the sheet determined in Ref.~\cite{King2012} should be compared with the extent $\ell$ of the cylindrical region.
For negative Gaussian curvatures, the edge stress that we predict would extend over a finite annulus~\cite{Timounay2021Sculpting} and the shape would be cylindrical in the inner region.

\begin{figure}
    \centering
    \includegraphics[width=.8\linewidth]{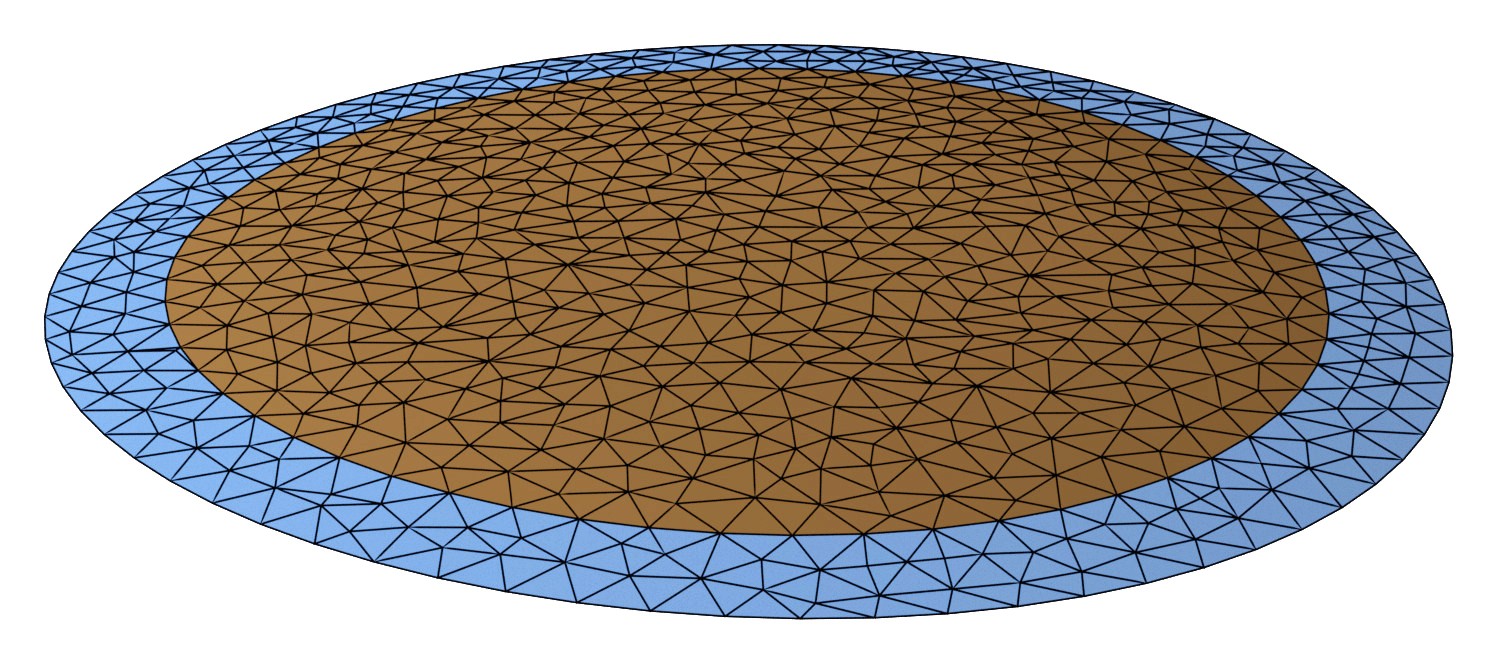}
    \caption{Rendering of the final configuration obtained by energy numerical energy minimization with the mesh represented as black segments. Parameters are $L=1.2$, $\kappa_1=0.1$, $\kappa_2=0.05$.
    \label{fig:se_blender}}
\end{figure}

\section{Numerical energy minimization with Surface Evolver}

We translate our membrane model into a computational model, using the ``Surface Evolver'' software \cite{Brakke1992}, where the system consists of a sheet attached to a liquid interface.
We use a triangulated surface (Fig.~\ref{fig:se_blender}) with a mesh size of $0.1$.
The liquid has surface tension $\gamma=1$ and the sheet is a disc of radius $W=1$.
The sheet has a bending modulus of zero to allow it to bend freely. In-plane sheet deformations are realized via an asymmetric elasticity: the compressive energy cost between points is zero, while points stretching past their initially prescribed distance is highly discouraged by using a large stretch modulus ($10^{12}$).
The boundary radius is fixed at $L=1.2$ and the boundary height and pressure are specified to achieve the desired curvature for each simulation. Starting at a stretching modulus of 10, we minimize the system energy using the conjugate gradient algorithm. We then increase the stretching modulus by a factor of $10^{0.001}$ and relax the system again (this factor yields 1000 evenly spaced steps for each order of magnitude of the stretch modulus).
This sequence is repeated until the stretch modulus is at the desired value.

\begin{figure}
    \centering
    \includegraphics{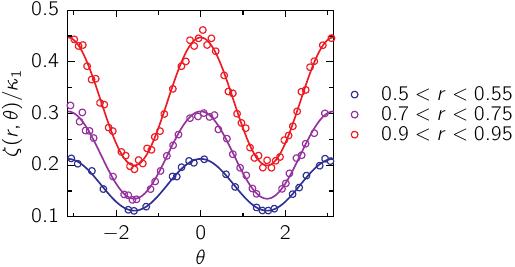}
    \caption{Vertical displacement as a function of the angle for different annuli, Surface Evolver simulations (circles) and fit with Eq.~(\ref{eq:vertical_displacement_form}) (solid lines). Parameters are $L=1.2$, $\kappa_1=0.1$, $\kappa_2=0.05$.
    \label{fig:fit_ci}}
\end{figure}

The height of an annulus around a radius $r$ can be fit with a height function of the form (Eq.~(3) in the main text)
\begin{equation}\label{eq:vertical_displacement_form}
 \zeta(r,\theta) = c_2(r)\cos(2\theta)+c_0(r),
\end{equation}
giving the values of $c_2(r)$ and $c_0(r)$ (Fig.~\ref{fig:fit_ci}).

\section{Experiment}

We form thin polymer films of thickness $49 < t < \qty{111}{\nano\meter}$ by spin coating dilute solutions of polystyrene ($M_\text{n} = 112.5$ k, $M_\text{w} = 118.5$ k, Polymer Source) in toluene ($99.9\%$, Fisher Scientific) on glass microscope slides. 
Film thickness may be varied by changing the polymer concentration and spinning speed; here we use polystyrene $2\%$ w/w or $1\%$ w/w, and we spin at $\qty{1000}{rpm}$ for $\qty{60}{\second}$. 
We cut the as-spun film into a disc of radius $W = \qty{1.7}{\milli\meter}$ and float it onto a dish of deionized water ($\gamma = \qty{72}{\milli\newton\per\meter}$), following Refs.~\cite{Huang2007,Timounay2020}. 
After the experiment, the film is captured on a silicon water and its thickness is measured using a white-light interferometer (Filmetrics F3). 

The mechanical properties of the film are set by its stretching and bending moduli, $Y=Et$ and $B=Et^3/[12(1-\nu^2)]$ respectively, and its Poisson's ratio $\nu=0.34$. 
Our parameters place us in the high bendability regime $\epsilon^{-1}=\gamma W^2/B \geq 4.7 \times 10^5$~\cite{King2012}: our films buckle under minute compression. 
At the same time, our films incur only small in-plane strains, as $\gamma/Y \leq 4.3 \times 10^{-4}$.


\begin{thebibliography}{26}%
\makeatletter
\providecommand \@ifxundefined [1]{%
 \@ifx{#1\undefined}
}%
\providecommand \@ifnum [1]{%
 \ifnum #1\expandafter \@firstoftwo
 \else \expandafter \@secondoftwo
 \fi
}%
\providecommand \@ifx [1]{%
 \ifx #1\expandafter \@firstoftwo
 \else \expandafter \@secondoftwo
 \fi
}%
\providecommand \natexlab [1]{#1}%
\providecommand \enquote  [1]{``#1''}%
\providecommand \bibnamefont  [1]{#1}%
\providecommand \bibfnamefont [1]{#1}%
\providecommand \citenamefont [1]{#1}%
\providecommand \href@noop [0]{\@secondoftwo}%
\providecommand \href [0]{\begingroup \@sanitize@url \@href}%
\providecommand \@href[1]{\@@startlink{#1}\@@href}%
\providecommand \@@href[1]{\endgroup#1\@@endlink}%
\providecommand \@sanitize@url [0]{\catcode `\\12\catcode `\$12\catcode
  `\&12\catcode `\#12\catcode `\^12\catcode `\_12\catcode `\%12\relax}%
\providecommand \@@startlink[1]{}%
\providecommand \@@endlink[0]{}%
\providecommand \url  [0]{\begingroup\@sanitize@url \@url }%
\providecommand \@url [1]{\endgroup\@href {#1}{\urlprefix }}%
\providecommand \urlprefix  [0]{URL }%
\providecommand \Eprint [0]{\href }%
\providecommand \doibase [0]{https://doi.org/}%
\providecommand \selectlanguage [0]{\@gobble}%
\providecommand \bibinfo  [0]{\@secondoftwo}%
\providecommand \bibfield  [0]{\@secondoftwo}%
\providecommand \translation [1]{[#1]}%
\providecommand \BibitemOpen [0]{}%
\providecommand \bibitemStop [0]{}%
\providecommand \bibitemNoStop [0]{.\EOS\space}%
\providecommand \EOS [0]{\spacefactor3000\relax}%
\providecommand \BibitemShut  [1]{\csname bibitem#1\endcsname}%
\let\auto@bib@innerbib\@empty
\bibitem [{\citenamefont {Marder}\ \emph {et~al.}(2007)\citenamefont {Marder},
  \citenamefont {Deegan},\ and\ \citenamefont {Sharon}}]{Marder2007Crumpling}%
  \BibitemOpen
  \bibfield  {author} {\bibinfo {author} {\bibfnamefont {M.}~\bibnamefont
  {Marder}}, \bibinfo {author} {\bibfnamefont {R.~D.}\ \bibnamefont {Deegan}},\
  and\ \bibinfo {author} {\bibfnamefont {E.}~\bibnamefont {Sharon}},\
  }\bibfield  {title} {\bibinfo {title} {{Crumpling, buckling, and cracking:
  Elasticity of thin sheets}},\ }\href {https://doi.org/10.1063/1.2711634}
  {\bibfield  {journal} {\bibinfo  {journal} {Physics Today}\ }\textbf
  {\bibinfo {volume} {60}},\ \bibinfo {pages} {33} (\bibinfo {year}
  {2007})}\BibitemShut {NoStop}%
\bibitem [{\citenamefont {Paulsen}(2019)}]{Paulsen2019}%
  \BibitemOpen
  \bibfield  {author} {\bibinfo {author} {\bibfnamefont {J.~D.}\ \bibnamefont
  {Paulsen}},\ }\bibfield  {title} {\bibinfo {title} {{Wrapping Liquids,
  Solids, and Gases in Thin Sheets}},\ }\href
  {https://doi.org/10.1146/annurev-conmatphys-031218-013533} {\bibfield
  {journal} {\bibinfo  {journal} {{Annual Review of Condensed Matter Physics}}\
  }\textbf {\bibinfo {volume} {10}},\ \bibinfo {pages} {431} (\bibinfo {year}
  {2019})}\BibitemShut {NoStop}%
\bibitem [{\citenamefont {Hure}\ \emph {et~al.}(2012)\citenamefont {Hure},
  \citenamefont {Roman},\ and\ \citenamefont {Bico}}]{Hure2012}%
  \BibitemOpen
  \bibfield  {author} {\bibinfo {author} {\bibfnamefont {J.}~\bibnamefont
  {Hure}}, \bibinfo {author} {\bibfnamefont {B.}~\bibnamefont {Roman}},\ and\
  \bibinfo {author} {\bibfnamefont {J.}~\bibnamefont {Bico}},\ }\bibfield
  {title} {\bibinfo {title} {{Stamping and Wrinkling of Elastic Plates}},\
  }\href {https://doi.org/10.1103/PhysRevLett.109.054302} {\bibfield  {journal}
  {\bibinfo  {journal} {{Phys. Rev. Lett.}}\ }\textbf {\bibinfo {volume}
  {109}},\ \bibinfo {pages} {054302} (\bibinfo {year} {2012})}\BibitemShut
  {NoStop}%
\bibitem [{\citenamefont {Davidovitch}\ \emph {et~al.}(2019)\citenamefont
  {Davidovitch}, \citenamefont {Sun},\ and\ \citenamefont
  {Grason}}]{Davidovitch2019}%
  \BibitemOpen
  \bibfield  {author} {\bibinfo {author} {\bibfnamefont {B.}~\bibnamefont
  {Davidovitch}}, \bibinfo {author} {\bibfnamefont {Y.}~\bibnamefont {Sun}},\
  and\ \bibinfo {author} {\bibfnamefont {G.~M.}\ \bibnamefont {Grason}},\
  }\bibfield  {title} {\bibinfo {title} {{Geometrically incompatible
  confinement of solids}},\ }\href {https://doi.org/10.1073/pnas.1815507116}
  {\bibfield  {journal} {\bibinfo  {journal} {{Proceedings of the National
  Academy of Sciences}}\ }\textbf {\bibinfo {volume} {116}},\ \bibinfo {pages}
  {1483} (\bibinfo {year} {2019})}\BibitemShut {NoStop}%
\bibitem [{\citenamefont {Yao}\ \emph {et~al.}(2013)\citenamefont {Yao},
  \citenamefont {Bowick}, \citenamefont {Ma},\ and\ \citenamefont
  {Sknepnek}}]{Yao2013}%
  \BibitemOpen
  \bibfield  {author} {\bibinfo {author} {\bibfnamefont {Z.}~\bibnamefont
  {Yao}}, \bibinfo {author} {\bibfnamefont {M.}~\bibnamefont {Bowick}},
  \bibinfo {author} {\bibfnamefont {X.}~\bibnamefont {Ma}},\ and\ \bibinfo
  {author} {\bibfnamefont {R.}~\bibnamefont {Sknepnek}},\ }\bibfield  {title}
  {\bibinfo {title} {{Planar sheets meet negative-curvature liquid
  interfaces}},\ }\href {https://doi.org/10.1209/0295-5075/101/44007}
  {\bibfield  {journal} {\bibinfo  {journal} {{EPL}}\ }\textbf {\bibinfo
  {volume} {101}},\ \bibinfo {pages} {44007} (\bibinfo {year}
  {2013})}\BibitemShut {NoStop}%
\bibitem [{\citenamefont {Bense}\ \emph {et~al.}(2020)\citenamefont {Bense},
  \citenamefont {Tani}, \citenamefont {Saint-Jean}, \citenamefont {Reyssat},
  \citenamefont {Roman},\ and\ \citenamefont {Bico}}]{Bense2020}%
  \BibitemOpen
  \bibfield  {author} {\bibinfo {author} {\bibfnamefont {H.}~\bibnamefont
  {Bense}}, \bibinfo {author} {\bibfnamefont {M.}~\bibnamefont {Tani}},
  \bibinfo {author} {\bibfnamefont {M.}~\bibnamefont {Saint-Jean}}, \bibinfo
  {author} {\bibfnamefont {E.}~\bibnamefont {Reyssat}}, \bibinfo {author}
  {\bibfnamefont {B.}~\bibnamefont {Roman}},\ and\ \bibinfo {author}
  {\bibfnamefont {J.}~\bibnamefont {Bico}},\ }\bibfield  {title} {\bibinfo
  {title} {{Elastocapillary adhesion of a soft cap on a rigid sphere}},\
  }\bibfield  {journal} {\bibinfo  {journal} {{Soft Matter}}\ }\href
  {https://doi.org/10.1039/C9SM02057H} {10.1039/C9SM02057H} (\bibinfo {year}
  {2020})\BibitemShut {NoStop}%
\bibitem [{\citenamefont {Box}\ \emph {et~al.}(2023)\citenamefont {Box},
  \citenamefont {Domino}, \citenamefont {Corvo}, \citenamefont {Adda-Bedia},
  \citenamefont {Démery}, \citenamefont {Vella},\ and\ \citenamefont
  {Davidovitch}}]{Box2023Delamination}%
  \BibitemOpen
  \bibfield  {author} {\bibinfo {author} {\bibfnamefont {F.}~\bibnamefont
  {Box}}, \bibinfo {author} {\bibfnamefont {L.}~\bibnamefont {Domino}},
  \bibinfo {author} {\bibfnamefont {T.~O.}\ \bibnamefont {Corvo}}, \bibinfo
  {author} {\bibfnamefont {M.}~\bibnamefont {Adda-Bedia}}, \bibinfo {author}
  {\bibfnamefont {V.}~\bibnamefont {Démery}}, \bibinfo {author} {\bibfnamefont
  {D.}~\bibnamefont {Vella}},\ and\ \bibinfo {author} {\bibfnamefont
  {B.}~\bibnamefont {Davidovitch}},\ }\bibfield  {title} {\bibinfo {title}
  {Delamination from an adhesive sphere: Curvature-induced dewetting versus
  buckling},\ }\href {https://doi.org/10.1073/pnas.2212290120} {\bibfield
  {journal} {\bibinfo  {journal} {Proceedings of the National Academy of
  Sciences}\ }\textbf {\bibinfo {volume} {120}},\ \bibinfo {pages}
  {e2212290120} (\bibinfo {year} {2023})}\BibitemShut {NoStop}%
\bibitem [{\citenamefont {Aharoni}\ \emph {et~al.}(2017)\citenamefont
  {Aharoni}, \citenamefont {Todorova}, \citenamefont {Albarrán}, \citenamefont
  {Goehring}, \citenamefont {Kamien},\ and\ \citenamefont
  {Katifori}}]{Aharoni2017}%
  \BibitemOpen
  \bibfield  {author} {\bibinfo {author} {\bibfnamefont {H.}~\bibnamefont
  {Aharoni}}, \bibinfo {author} {\bibfnamefont {D.~V.}\ \bibnamefont
  {Todorova}}, \bibinfo {author} {\bibfnamefont {O.}~\bibnamefont {Albarrán}},
  \bibinfo {author} {\bibfnamefont {L.}~\bibnamefont {Goehring}}, \bibinfo
  {author} {\bibfnamefont {R.~D.}\ \bibnamefont {Kamien}},\ and\ \bibinfo
  {author} {\bibfnamefont {E.}~\bibnamefont {Katifori}},\ }\bibfield  {title}
  {\bibinfo {title} {The smectic order of wrinkles},\ }\href
  {https://doi.org/10.1038/ncomms15809} {\bibfield  {journal} {\bibinfo
  {journal} {Nature Communications}\ }\textbf {\bibinfo {volume} {8}},\
  \bibinfo {pages} {15809} (\bibinfo {year} {2017})}\BibitemShut {NoStop}%
\bibitem [{\citenamefont {Tobasco}(2021)}]{Tobasco2021CurvatureDriven}%
  \BibitemOpen
  \bibfield  {author} {\bibinfo {author} {\bibfnamefont {I.}~\bibnamefont
  {Tobasco}},\ }\bibfield  {title} {\bibinfo {title} {Curvature-driven
  wrinkling of thin elastic shells},\ }\href
  {https://doi.org/10.1007/s00205-020-01566-8} {\bibfield  {journal} {\bibinfo
  {journal} {Archive for Rational Mechanics and Analysis}\ }\textbf {\bibinfo
  {volume} {239}},\ \bibinfo {pages} {1211} (\bibinfo {year}
  {2021})}\BibitemShut {NoStop}%
\bibitem [{\citenamefont {Tobasco}\ \emph {et~al.}(2022)\citenamefont
  {Tobasco}, \citenamefont {Timounay}, \citenamefont {Todorova}, \citenamefont
  {Leggat}, \citenamefont {Paulsen},\ and\ \citenamefont
  {Katifori}}]{Tobasco2022Exact}%
  \BibitemOpen
  \bibfield  {author} {\bibinfo {author} {\bibfnamefont {I.}~\bibnamefont
  {Tobasco}}, \bibinfo {author} {\bibfnamefont {Y.}~\bibnamefont {Timounay}},
  \bibinfo {author} {\bibfnamefont {D.}~\bibnamefont {Todorova}}, \bibinfo
  {author} {\bibfnamefont {G.~C.}\ \bibnamefont {Leggat}}, \bibinfo {author}
  {\bibfnamefont {J.~D.}\ \bibnamefont {Paulsen}},\ and\ \bibinfo {author}
  {\bibfnamefont {E.}~\bibnamefont {Katifori}},\ }\bibfield  {title} {\bibinfo
  {title} {Exact solutions for the wrinkle patterns of confined elastic
  shells},\ }\href {https://doi.org/10.1038/s41567-022-01672-2} {\bibfield
  {journal} {\bibinfo  {journal} {Nature Physics}\ } (\bibinfo {year}
  {2022})}\BibitemShut {NoStop}%
\bibitem [{\citenamefont {Timounay}\ \emph {et~al.}(2021)\citenamefont
  {Timounay}, \citenamefont {Hartwell}, \citenamefont {He}, \citenamefont
  {King}, \citenamefont {Murphy}, \citenamefont {D\'emery},\ and\ \citenamefont
  {Paulsen}}]{Timounay2021Sculpting}%
  \BibitemOpen
  \bibfield  {author} {\bibinfo {author} {\bibfnamefont {Y.}~\bibnamefont
  {Timounay}}, \bibinfo {author} {\bibfnamefont {A.~R.}\ \bibnamefont
  {Hartwell}}, \bibinfo {author} {\bibfnamefont {M.}~\bibnamefont {He}},
  \bibinfo {author} {\bibfnamefont {D.~E.}\ \bibnamefont {King}}, \bibinfo
  {author} {\bibfnamefont {L.~K.}\ \bibnamefont {Murphy}}, \bibinfo {author}
  {\bibfnamefont {V.}~\bibnamefont {D\'emery}},\ and\ \bibinfo {author}
  {\bibfnamefont {J.~D.}\ \bibnamefont {Paulsen}},\ }\bibfield  {title}
  {\bibinfo {title} {Sculpting liquids with ultrathin shells},\ }\href
  {https://doi.org/10.1103/PhysRevLett.127.108002} {\bibfield  {journal}
  {\bibinfo  {journal} {Phys. Rev. Lett.}\ }\textbf {\bibinfo {volume} {127}},\
  \bibinfo {pages} {108002} (\bibinfo {year} {2021})}\BibitemShut {NoStop}%
\bibitem [{\citenamefont {King}\ \emph {et~al.}(2012)\citenamefont {King},
  \citenamefont {Schroll}, \citenamefont {Davidovitch},\ and\ \citenamefont
  {Menon}}]{King2012}%
  \BibitemOpen
  \bibfield  {author} {\bibinfo {author} {\bibfnamefont {H.}~\bibnamefont
  {King}}, \bibinfo {author} {\bibfnamefont {R.~D.}\ \bibnamefont {Schroll}},
  \bibinfo {author} {\bibfnamefont {B.}~\bibnamefont {Davidovitch}},\ and\
  \bibinfo {author} {\bibfnamefont {N.}~\bibnamefont {Menon}},\ }\bibfield
  {title} {\bibinfo {title} {{Elastic sheet on a liquid drop reveals wrinkling
  and crumpling as distinct symmetry-breaking instabilities}},\ }\href
  {https://doi.org/10.1073/pnas.1201201109} {\bibfield  {journal} {\bibinfo
  {journal} {{Proc. Natl. Acad. Sci. U.S.A.}}\ }\textbf {\bibinfo {volume}
  {109}},\ \bibinfo {pages} {9716} (\bibinfo {year} {2012})}\BibitemShut
  {NoStop}%
\bibitem [{\citenamefont {Paulsen}\ \emph {et~al.}(2015)\citenamefont
  {Paulsen}, \citenamefont {Démery}, \citenamefont {Santangelo}, \citenamefont
  {Russell}, \citenamefont {Davidovitch},\ and\ \citenamefont
  {Menon}}]{Paulsen2015}%
  \BibitemOpen
  \bibfield  {author} {\bibinfo {author} {\bibfnamefont {J.~D.}\ \bibnamefont
  {Paulsen}}, \bibinfo {author} {\bibfnamefont {V.}~\bibnamefont {Démery}},
  \bibinfo {author} {\bibfnamefont {C.~D.}\ \bibnamefont {Santangelo}},
  \bibinfo {author} {\bibfnamefont {T.~P.}\ \bibnamefont {Russell}}, \bibinfo
  {author} {\bibfnamefont {B.}~\bibnamefont {Davidovitch}},\ and\ \bibinfo
  {author} {\bibfnamefont {N.}~\bibnamefont {Menon}},\ }\bibfield  {title}
  {\bibinfo {title} {{Optimal wrapping of liquid droplets with ultrathin
  sheets}},\ }\href {{http://dx.doi.org/10.1038/nmat4397}} {\bibfield
  {journal} {\bibinfo  {journal} {{Nat Mater}}\ }\textbf {\bibinfo {volume}
  {14}},\ \bibinfo {pages} {1206} (\bibinfo {year} {2015})},\ \bibinfo {note}
  {{Letter}}\BibitemShut {NoStop}%
\bibitem [{\citenamefont {Bae}\ \emph {et~al.}(2015)\citenamefont {Bae},
  \citenamefont {Ouchi},\ and\ \citenamefont {Hayward}}]{Bae2015Measuring}%
  \BibitemOpen
  \bibfield  {author} {\bibinfo {author} {\bibfnamefont {J.}~\bibnamefont
  {Bae}}, \bibinfo {author} {\bibfnamefont {T.}~\bibnamefont {Ouchi}},\ and\
  \bibinfo {author} {\bibfnamefont {R.~C.}\ \bibnamefont {Hayward}},\
  }\bibfield  {title} {\bibinfo {title} {Measuring the elastic modulus of thin
  polymer sheets by elastocapillary bending},\ }\href
  {https://doi.org/10.1021/acsami.5b02567} {\bibfield  {journal} {\bibinfo
  {journal} {ACS Appl. Mater. Interfaces}\ }\textbf {\bibinfo {volume} {7}},\
  \bibinfo {pages} {14734} (\bibinfo {year} {2015})}\BibitemShut {NoStop}%
\bibitem [{\citenamefont {Kumar}\ \emph {et~al.}(2018)\citenamefont {Kumar},
  \citenamefont {Paulsen}, \citenamefont {Russell},\ and\ \citenamefont
  {Menon}}]{Kumar2018}%
  \BibitemOpen
  \bibfield  {author} {\bibinfo {author} {\bibfnamefont {D.}~\bibnamefont
  {Kumar}}, \bibinfo {author} {\bibfnamefont {J.~D.}\ \bibnamefont {Paulsen}},
  \bibinfo {author} {\bibfnamefont {T.~P.}\ \bibnamefont {Russell}},\ and\
  \bibinfo {author} {\bibfnamefont {N.}~\bibnamefont {Menon}},\ }\bibfield
  {title} {\bibinfo {title} {{Wrapping with a splash: High-speed encapsulation
  with ultrathin sheets}},\ }\href {https://doi.org/10.1126/science.aao1290}
  {\bibfield  {journal} {\bibinfo  {journal} {{Science}}\ }\textbf {\bibinfo
  {volume} {359}},\ \bibinfo {pages} {775} (\bibinfo {year}
  {2018})}\BibitemShut {NoStop}%
\bibitem [{\citenamefont {Hohlfeld}\ and\ \citenamefont
  {Davidovitch}(2015)}]{Hohlfeld2015}%
  \BibitemOpen
  \bibfield  {author} {\bibinfo {author} {\bibfnamefont {E.}~\bibnamefont
  {Hohlfeld}}\ and\ \bibinfo {author} {\bibfnamefont {B.}~\bibnamefont
  {Davidovitch}},\ }\bibfield  {title} {\bibinfo {title} {{Sheet on a
  deformable sphere: Wrinkle patterns suppress curvature-induced
  delamination}},\ }\href {https://doi.org/10.1103/PhysRevE.91.012407}
  {\bibfield  {journal} {\bibinfo  {journal} {{Phys. Rev. E}}\ }\textbf
  {\bibinfo {volume} {91}},\ \bibinfo {pages} {012407} (\bibinfo {year}
  {2015})}\BibitemShut {NoStop}%
\bibitem [{\citenamefont {Paulsen}\ \emph {et~al.}(2017)\citenamefont
  {Paulsen}, \citenamefont {Démery}, \citenamefont {Toga}, \citenamefont
  {Qiu}, \citenamefont {Russell}, \citenamefont {Davidovitch},\ and\
  \citenamefont {Menon}}]{Paulsen2017}%
  \BibitemOpen
  \bibfield  {author} {\bibinfo {author} {\bibfnamefont {J.~D.}\ \bibnamefont
  {Paulsen}}, \bibinfo {author} {\bibfnamefont {V.}~\bibnamefont {Démery}},
  \bibinfo {author} {\bibfnamefont {K.~B.}\ \bibnamefont {Toga}}, \bibinfo
  {author} {\bibfnamefont {Z.}~\bibnamefont {Qiu}}, \bibinfo {author}
  {\bibfnamefont {T.~P.}\ \bibnamefont {Russell}}, \bibinfo {author}
  {\bibfnamefont {B.}~\bibnamefont {Davidovitch}},\ and\ \bibinfo {author}
  {\bibfnamefont {N.}~\bibnamefont {Menon}},\ }\bibfield  {title} {\bibinfo
  {title} {{Geometry-Driven Folding of a Floating Annular Sheet}},\ }\href
  {https://doi.org/10.1103/PhysRevLett.118.048004} {\bibfield  {journal}
  {\bibinfo  {journal} {{Phys. Rev. Lett.}}\ }\textbf {\bibinfo {volume}
  {118}},\ \bibinfo {pages} {048004} (\bibinfo {year} {2017})}\BibitemShut
  {NoStop}%
\bibitem [{SM()}]{SM}%
  \BibitemOpen
  \href@noop {} {\bibinfo {title} {See supplemental material at [url will be
  inserted by publisher] for supplementary figures, methods, and
  calculations.}}\BibitemShut {Stop}%
\bibitem [{\citenamefont {Pak}\ and\ \citenamefont {Schlenker}(2010)}]{Pak10}%
  \BibitemOpen
  \bibfield  {author} {\bibinfo {author} {\bibfnamefont {I.}~\bibnamefont
  {Pak}}\ and\ \bibinfo {author} {\bibfnamefont {J.-M.}\ \bibnamefont
  {Schlenker}},\ }\bibfield  {title} {\bibinfo {title} {Profiles of inflated
  surfaces},\ }\href@noop {} {\bibfield  {journal} {\bibinfo  {journal}
  {Journal of Nonlinear Mathematical Physics}\ }\textbf {\bibinfo {volume}
  {17}},\ \bibinfo {pages} {145} (\bibinfo {year} {2010})}\BibitemShut
  {NoStop}%
\bibitem [{\citenamefont {Brakke}(1992)}]{Brakke1992}%
  \BibitemOpen
  \bibfield  {author} {\bibinfo {author} {\bibfnamefont {K.~A.}\ \bibnamefont
  {Brakke}},\ }\bibfield  {title} {\bibinfo {title} {{The Surface Evolver}},\
  }\href {https://doi.org/10.1080/10586458.1992.10504253} {\bibfield  {journal}
  {\bibinfo  {journal} {{Experimental Mathematics}}\ }\textbf {\bibinfo
  {volume} {1}},\ \bibinfo {pages} {141} (\bibinfo {year} {1992})}\BibitemShut
  {NoStop}%
\bibitem [{\citenamefont {Taylor}(1919)}]{Taylor1919}%
  \BibitemOpen
  \bibfield  {author} {\bibinfo {author} {\bibfnamefont {G.}~\bibnamefont
  {Taylor}},\ }\bibfield  {title} {\bibinfo {title} {{On the shape of
  parachutes}},\ }\href@noop {} {\bibfield  {journal} {\bibinfo  {journal}
  {{Advisory Committee for Aeronautics}}\ } (\bibinfo {year}
  {1919})}\BibitemShut {NoStop}%
\bibitem [{\citenamefont {Knops}\ and\ \citenamefont
  {Payne}(1971)}]{Knops1971Uniqueness}%
  \BibitemOpen
  \bibfield  {author} {\bibinfo {author} {\bibfnamefont {R.~J.}\ \bibnamefont
  {Knops}}\ and\ \bibinfo {author} {\bibfnamefont {L.~E.}\ \bibnamefont
  {Payne}},\ }\bibfield  {title} {\bibinfo {title} {Uniqueness theorems in
  linear elasticity},\ }\href@noop {} {\bibfield  {journal} {\bibinfo
  {journal} {Springer Tracts in Natural Phylosophy}\ }\textbf {\bibinfo
  {volume} {19}} (\bibinfo {year} {1971})}\BibitemShut {NoStop}%
\bibitem [{\citenamefont {Huang}\ \emph {et~al.}(2007)\citenamefont {Huang},
  \citenamefont {Juszkiewicz}, \citenamefont {de~Jeu}, \citenamefont {Cerda},
  \citenamefont {Emrick}, \citenamefont {Menon},\ and\ \citenamefont
  {Russell}}]{Huang2007}%
  \BibitemOpen
  \bibfield  {author} {\bibinfo {author} {\bibfnamefont {J.}~\bibnamefont
  {Huang}}, \bibinfo {author} {\bibfnamefont {M.}~\bibnamefont {Juszkiewicz}},
  \bibinfo {author} {\bibfnamefont {W.~H.}\ \bibnamefont {de~Jeu}}, \bibinfo
  {author} {\bibfnamefont {E.}~\bibnamefont {Cerda}}, \bibinfo {author}
  {\bibfnamefont {T.}~\bibnamefont {Emrick}}, \bibinfo {author} {\bibfnamefont
  {N.}~\bibnamefont {Menon}},\ and\ \bibinfo {author} {\bibfnamefont {T.~P.}\
  \bibnamefont {Russell}},\ }\bibfield  {title} {\bibinfo {title} {{Capillary
  Wrinkling of Floating Thin Polymer Films}},\ }\href
  {https://doi.org/10.1126/science.1144616} {\bibfield  {journal} {\bibinfo
  {journal} {{Science}}\ }\textbf {\bibinfo {volume} {317}},\ \bibinfo {pages}
  {650} (\bibinfo {year} {2007})}\BibitemShut {NoStop}%
\bibitem [{\citenamefont {Timounay}\ \emph {et~al.}(2020)\citenamefont
  {Timounay}, \citenamefont {De}, \citenamefont {Stelzel}, \citenamefont
  {Schrecengost}, \citenamefont {Ripp},\ and\ \citenamefont
  {Paulsen}}]{Timounay2020}%
  \BibitemOpen
  \bibfield  {author} {\bibinfo {author} {\bibfnamefont {Y.}~\bibnamefont
  {Timounay}}, \bibinfo {author} {\bibfnamefont {R.}~\bibnamefont {De}},
  \bibinfo {author} {\bibfnamefont {J.~L.}\ \bibnamefont {Stelzel}}, \bibinfo
  {author} {\bibfnamefont {Z.~S.}\ \bibnamefont {Schrecengost}}, \bibinfo
  {author} {\bibfnamefont {M.~M.}\ \bibnamefont {Ripp}},\ and\ \bibinfo
  {author} {\bibfnamefont {J.~D.}\ \bibnamefont {Paulsen}},\ }\bibfield
  {title} {\bibinfo {title} {{Crumples as a Generic Stress-Focusing Instability
  in Confined Sheets}},\ }\href {https://doi.org/10.1103/PhysRevX.10.021008}
  {\bibfield  {journal} {\bibinfo  {journal} {{Phys. Rev. X}}\ }\textbf
  {\bibinfo {volume} {10}},\ \bibinfo {pages} {021008} (\bibinfo {year}
  {2020})}\BibitemShut {NoStop}%
\bibitem [{\citenamefont {Han}\ \emph {et~al.}(2006)\citenamefont {Han},
  \citenamefont {Hong},\ and\ \citenamefont {Hong}}]{Han2006Isometric}%
  \BibitemOpen
  \bibfield  {author} {\bibinfo {author} {\bibfnamefont {Q.}~\bibnamefont
  {Han}}, \bibinfo {author} {\bibfnamefont {J.-X.}\ \bibnamefont {Hong}},\ and\
  \bibinfo {author} {\bibfnamefont {J.}~\bibnamefont {Hong}},\ }\href@noop {}
  {\emph {\bibinfo {title} {Isometric embedding of Riemannian manifolds in
  Euclidean spaces}}},\ Vol.~\bibinfo {volume} {13}\ (\bibinfo  {publisher}
  {American Mathematical Soc.},\ \bibinfo {year} {2006})\BibitemShut {NoStop}%
\bibitem [{\citenamefont {Py}\ \emph {et~al.}(2007)\citenamefont {Py},
  \citenamefont {Reverdy}, \citenamefont {Doppler}, \citenamefont {Bico},
  \citenamefont {Roman},\ and\ \citenamefont {Baroud}}]{Py2007}%
  \BibitemOpen
  \bibfield  {author} {\bibinfo {author} {\bibfnamefont {C.}~\bibnamefont
  {Py}}, \bibinfo {author} {\bibfnamefont {P.}~\bibnamefont {Reverdy}},
  \bibinfo {author} {\bibfnamefont {L.}~\bibnamefont {Doppler}}, \bibinfo
  {author} {\bibfnamefont {J.}~\bibnamefont {Bico}}, \bibinfo {author}
  {\bibfnamefont {B.}~\bibnamefont {Roman}},\ and\ \bibinfo {author}
  {\bibfnamefont {C.~N.}\ \bibnamefont {Baroud}},\ }\bibfield  {title}
  {\bibinfo {title} {{Capillary Origami: Spontaneous Wrapping of a Droplet with
  an Elastic Sheet}},\ }\href {https://doi.org/10.1103/PhysRevLett.98.156103}
  {\bibfield  {journal} {\bibinfo  {journal} {{Phys. Rev. Lett.}}\ }\textbf
  {\bibinfo {volume} {98}},\ \bibinfo {pages} {156103} (\bibinfo {year}
  {2007})}\BibitemShut {NoStop}%
\end{thebibliography}
%

\end{document}